# Critical Current Longitudinal and Transverse Strain Sensitivities of High $J_C$ Nb$_3$Sn Conductors

Jun Lu, Ke Han, Robert P. Walsh, Todd Atkins and Scott T. Bole

*Abstract*—Characterizing critical current $I_C$ of Nb$_3$Sn strands as function of a strain is very important for large high field superconducting magnet applications such as the superconducting outsert coil of the series-connected hybrid at the NHMFL and the ITER magnets. Apparatuses for measuring $I_C$ versus longitudinal strain and transverse stress have been developed and used at the NHMFL. We have characterized the $I_C$ strain sensitivities of a few candidate strands for the series-connected-hybrid. In addition, $I_C$ irreversibility strains are measured for the recently developed ITER high $J_C$ strands. The different strain sensitivities for different strands are discussed.

## I. INTRODUCTION

THE series-connected hybrid magnet (SCH) under construction at the National High Magnetic Field Laboratory, USA (NHMFL) has superconducting outsert coils that use cable-in-conduit conductor (CICC) with high $J_C$ (~ 2000 A/mm$^2$ non-Cu) Nb$_3$Sn strands. It is well known that $I_C$ of Nb$_3$Sn is sensitive to longitudinal and transverse strains especially at high fields [1], [2]. In an energized CICC, the strains in strands come from the differential thermal contraction, longitudinal hoop stress and localized transverse compression and bending due to large transverse Lorentz force [3]. The SCH outsert coils operate at a peak field of ~14 T and a current of 20 kA. So the maximum transverse Lorentz force is ~280 kN/m. For the currently designed conduit size, the maximum stress experienced by a strand is ~ 16 MPa, comparable with that of ITER conductors. Under this large transverse load, the localized transverse compressive stress at the strands cross-overs and the bending strain in between may cause significant reduction of $I_C$. Although optimizing the CICC cable pattern and the void fraction might mitigate the problem, use of less strain sensitive strands is always desirable. Unfortunately $I_C$ strain sensitivity is strand design specific, and the correlation between $I_C$ strain sensitivity and strand design is only beginning to be understood [4]. Therefore at present the experimental measurement is the only way to provide reliable strand $I_C$ strain characteristics which are crucial for a reliable magnet design and successful construction.

In other words, the strand $I_C$ strain characterization of strands with different designs is a necessary part of the strand selection process for SCH where a combination of high $I_C$ and low strain sensitivity and low AC loss is desired. Meanwhile, the experimental strand $I_C$ strain data will be used as inputs to calculate the CICC performance under operational conditions based on CICC strain models developed recently [3], [5], [6]. Strand $I_C$ reduction with bending strain is found to be the primary reason for the CICC $I_C$ reduction under transverse Lorentz force in ITER model coils. $I_C$ vs. bending strain is an important parameter in these models and a direct measurement has been performed [3]. Nevertheless, to the first order approximation, the $I_C$ vs. longitudinal strain can be used to estimate the $I_C$ under bending strain [7].

Moreover, under large localized strain, e.g. the outer rim of a strand under large bending, filament breakage occurs and $I_C$ reduction becomes irreversible. This causes the CICC $I_C$ degradation especially under cyclic loading. The strain when this happens is called irreversibility strain. Obviously, the larger the irreversibility strain, the less $I_C$ degradation under large bending strain. Therefore the irreversibility strain is an indicator of strand performance under large bending strain.

Finally, for the fundamental understanding of A-15 material's $I_C$-strain properties, it is necessary to have sufficient large body of $I_C$-strain data in both longitudinal and transverse directions for a variety of strands.

In this work, we characterized $I_C$ vs. longitudinal stress and transverse strain for the SCH candidate strands. In addition, the irreversibility strain is measured for four recently developed ITER TF high $J_C$ strands.

## II. EXPERIMENT

An apparatus for $I_C$ longitudinal strain measurements was designed and built taking advantage of the large 195 mm bore 0-20 T resistive magnet at the NHMFL. A simple mechanism is used to pull a ~100 mm long straight strand soldered on a backing plate (Fig. 1). More details on this device can be found in [8]. In this device, the dog-bone-shaped backing plate is used for better mechanical, electrical and thermal stability. The choice of the backing plate material is important. Ideally, it has large elastic range at liquid helium temperature, and has good electrical and thermal conductivity, and it is easy to solder the strands on to. We chose half hard and annealed CuBe (alloy 25) for most of our tests, although some earlier results were obtained using other backing plate materials.

To measure $I_C$ as a function of transverse stress, another

Manuscript received August 28, 2007. This work was performed at the National High Magnetic Field Laboratory which is supported by NSF under cooperative agreement No. DMR-0084173 and the State of Florida. The financial support of NSF SCH program and ITER US team is also acknowledged. Authors are with the National High Magnetic Field Laboratory, Tallahassee, FL 32310 USA (corresponding author J. Lu. Phone: 850-644-1678; fax: 850-644-0867; e-mail: junlu@magnet.fsu.edu).

device was recently designed and built as shown in Fig. 2. An air cylinder at the top of the probe pulls a puck (not shown) via a steel cable, so the puck's top surface applies transverse stresses on a ~70 mm long U-shaped Nb$_3$Sn strand lying on the bottom of the sample holder as shown in the close-in schematic in Fig. 2. The applied load is calibrated against the air pressure of the air cylinder. The stress is calculated using the applied force divided by the projected area, i.e. the product of the wire diameter and its length. The stress uniformity on the sample is ensured by using a half circle Nb$_3$Sn wire which is electrically insulated to complete a full Nb$_3$Sn circle in the sampling region. However, although the original design has a shallow groove on the sample holder to mechanically support the sample against Lorentz force, for the data presented here, we soldered the sample down on a CuAg backing plate for better mechanical support as well as electrical and thermal stability. The disadvantage of soldering wire to a substrate is that small uneven solder will results in some stress inhomogeneity. In addition, the thermal contraction difference between the backing plate and the sample introduces an additional thermal strain in both longitudinal and the orthogonal transverse directions. Therefore corrections are needed in data analysis to take these effects into account. Modifications in the design are being made to deal with these issues.

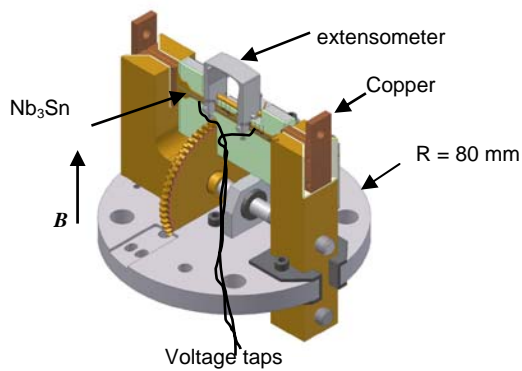

Fig. 1. The device used to measure $I_C$ vs. longitudinal strain.

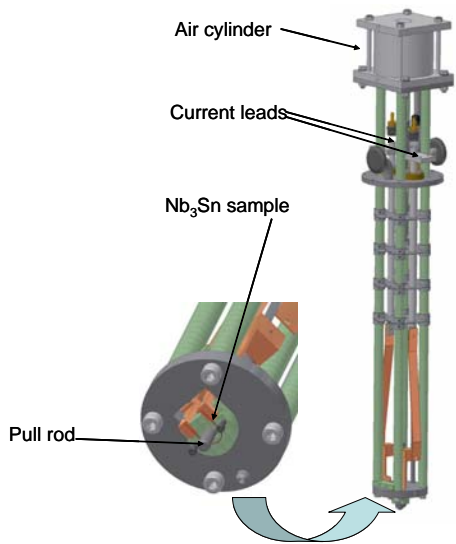

Fig. 2. The device used to measure $I_C$ vs. transverse stress.

## III. RESULTS AND DISCUSSIONS

### A. SCH strands

The SCH candidate strands are 0.6 mm in diameter and the specified $J_C$ is $> 2000$ A/mm$^2$ ($I_C > 247$ A) at 12 T 4.2 K [9]. $I_C$ vs. longitudinal strain is measured for all SCH candidate wires by using criterion of 1 µV/cm in at least two different fields each. So $B_{c2}^*$ vs. longitudinal strain can also be characterized. The Summers scaling law [10] is used to fit $I_C(B,\varepsilon)$ experimental data, where the critical temperature at zero strain and zero field $T_{cm0}$ is assumed to be 17.5 K. For detailed fitting procedures, see reference [8]. Fig. 3 shows the $I_C$ vs. intrinsic longitudinal strain curves for a few SCH candidate wires. The curves are calculated for 14 T and 4.2 K, which is the peak field in SCH, using the Summers fitting parameters listed in Table 1.

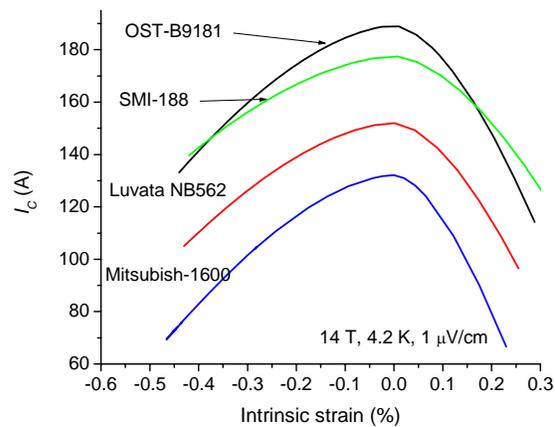

Fig. 3 $I_C$ vs. longitudinal strain for SCH candidate wires calculated for 14 T and 4.2 K using scaling parameters from Table 1.

TABLE 1 FITTING PARAMETERS OF $I_C$-$\varepsilon$ CURVES

| Strand | a+ | a- | $B_{c20m}$(T) | $T_{c0m}$(K) | $C_0$ (AT$^{1/2}$/mm$^2$) |
|---|---|---|---|---|---|
| Mitsubishi-1600 | 1150 | 4000 | 26.2 | 17.5 | 22300 |
| SMI-188 | 750 | 1850 | 30.1 | 17.5 | 22300 |
| Luvata-NB562 | 900 | 2600 | 26.6 | 17.5 | 25800 |
| OST-B9181 | 900 | 2500 | 27.8 | 17.5 | 28400 |

It is evident that different strands have different $I_C$ at zero intrinsic strain. However, it should be noted that in Fig. 3 $I_C$ is plotted for SCH engineering purposes. Since the non-Cu fraction is slightly different for each wire, a $J_C$ vs. strain plot looks slightly different from Fig. 3. From the Summers scaling law, the greater a+ and a-, the more $I_C$ strain sensitive. The higher the $B_{c20m}$ and $T_{c0m}$ the less is $I_C$ strain sensitive.

The $B_{c20m}$ values in table 1 are fitting parameters essentially obtained by Kramer extrapolations. This $B_{c20m}$ is used to do field extrapolation and give satisfactory results in fields much before $B_{c20}$, However, it should be noted that the directly measured $B_{c20}$ values may be significantly lower due to the significant deviation from the Kramer law near $B_{c20}$ [11]. On the other hand, due to the variation of tin content in $Nb_3Sn$ in a strand, there is a $T_C$ distribution. C. Senatore [12] measured strand $T_C$ distributions using specific heat method and obtained distributions with peaks at ~17.2 K and ~17.8 K for an internal-tin and a PIT strands respectively. For our SCH application near 4.2 K that is much below $T_C$, the effect of small difference in $T_C$ and $T_C$ distributions is much less significant. So we assume $T_C$ = 17.5 K for all our samples.

From table 1, OST-B9181 strand has the highest $I_C$ and moderate strain sensitivity as indicated by $a+$, $a-$ and $B_{c20m}$ values. SMI-188 has slightly lower $I_C$ but is less strain sensitive. So that at larger strain (>0.2% or <-0.35%), the $I_C$ of SMI-188 is greater than that of OST-B9181. Since in an energized CICC, each strand experiences a distributed strain, e.g. bending strain, it is necessary to evaluate integrated strand $I_C$ over a range of strains. Of course, for this evaluation a CICC strain model [3], [5], [6] is needed.

Fig. 4 is the $I_C$ vs. transverse stress of the SMI-188 wire. Similar to the results reported by other groups on different $Nb_3Sn$ wires [2], [13], [14], [17], the $I_C$ decreases monotonically with stress. For comparison, a calculated $I_C$ vs. longitudinal stress of the same wire based on Summers fitting parameters of table 1 is plotted with $I_C$-transverse stress in Fig. 5. The longitudinal compressive stress is obtained from the strain value and a $Nb_3Sn$ elastic modulus at 4 K of 65 GPa [15]. Since this wire has Ta addition, it may have higher modulus. To compensate the difference in backing materials of the $I_C$ transverse and longitudinal stress data, the $I_C$ longitudinal stress curve is shifted in stress axis in Fig. 5 so that the $I_C$ at zero longitudinal stress is the same as that at zero transverse stress.

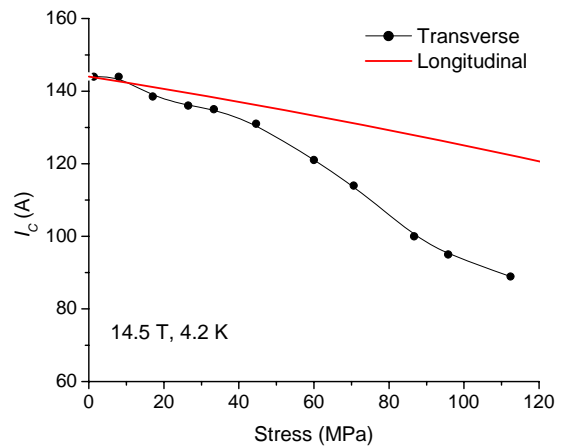

Fig. 5. Comparison of $I_C$ vs. transverse and longitudinal stress for SMI-188 PIT strand.

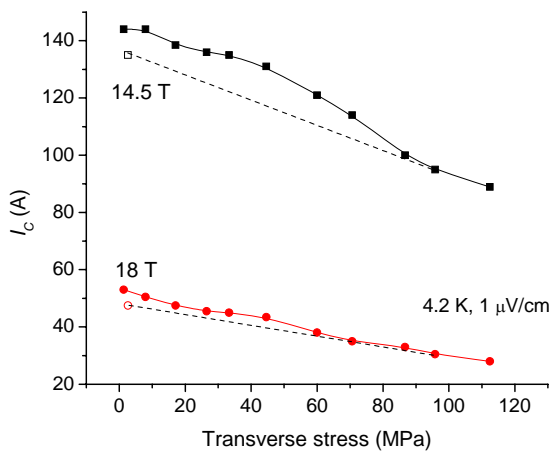

Fig. 4. $I_C$ vs. transverse stress for SMI-188 PIT strand at 14.5 T and 18 T.

The difference between the $I_C$ transverse and longituduinal stress curves in Fig. 5 is prominent, but less than what is reported for bronze route wire in [2] partly due to the use of a larger elastic modulus of 165 GPa in the conversion of longitudinal strain to stress in [2].

### B. ITER strands

Recently, high $J_C$ ITER TF strands are manufactured by Luvata and OST. The non-Cu $J_C$ at 4.2 K and 12 T is ~1000 A/mm$^2$. Previously, ITER conductor testing results showed that the significant $I_C$ degradation in CICC made from high $J_C$ strands eats away the benefit of high $J_C$ strand [16]. The $I_C$ degradation under cyclic loading indicates the irreversible $Nb_3Sn$ filament damage. Therefore it is desirable to use high $J_C$ strands that can withstand large strain without filament breaking, i.e. that have large irreversibility strains. The longitudinal irreversibility strains are measured. In these measurements, $I_C$ is initially measured with increasing strain. Once the $\varepsilon_{max}$ is reached, an additional $I_C$ is measured after a ~0.1-0.2% reduced strain for each strain increment. The irreversibility strain is the difference between $\varepsilon_{max}$ and $\varepsilon_{irr}$ after which $I_C$ is not reversible. We have tested four ITER high $J_C$ strands, two from the Luvata and two from the OST.

The strands are heat treated in vacuum sealed quartz tubes with following schedules.

210 °C/50 h + 340 °C/25 h + 450 °C/25 h + 575 °C/100 h + 650 °C/(100 h for OST, 200 h for Luvata),  with a ramp rate of 50 °C/h.

This heat treatment schedule is modified based on the ITER schedule where the ramp rate is 5 °C/h.

Results of $I_C$ vs. longitudinal strain at 18 T for two wires are shown in Fig. 6(a) and 6(b). Since $I_C$ is more strain sensitive at high field, the test was done at 18 T. Due to the lower n values at high field, the error in $I_C$ determination may be greater than that measured in lower fields. The Summers scaling law fittings indicate insignificant difference in $a+$, $a-$ and $B_{c20m}$ values between the two wires. For both tests the $\varepsilon_{max}$ is at ~0.35% consistent with the thermal contraction difference between the $Nb_3Sn$ wire and the CuBe backing plate as described in [8]. The OST-9355 has high $I_C$ at 18 T, whereas Luvata-8406 has higher irreversibility strain. Table 2 listed the irreversibility strains for all four samples measured. Two

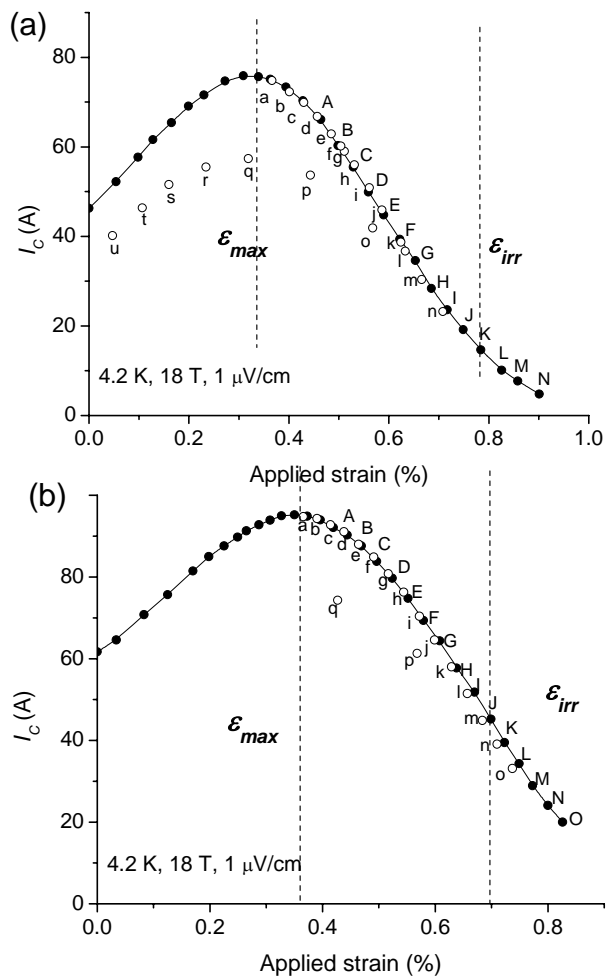

Fig. 6. $I_C$ vs. longitudinal strain for (a) ITER Luvata-8406 and (b) ITER OST-9355. CuBe backing plates were used.

Luvata samples have consistently higher irreversibility strains. It is interesting to note that with similar Summers fitting parameters, the irreversibility strains of the Luvata and the OST wires are appreciably different. It seems that the value of the irreversibility strain is independent of wires strain sensitivity. The microstructure analysis of these wires is underway and hopefully will give us more insights of this important property.

TABLE 2. IRREVERSIBILITY STRAINS OF ITER WIRES

|  | OST |  | Luvata |  |
|---|---|---|---|---|
| Wire ID | 9707 | 9355 | 8401 | 8406 |
| Diameter (mm) | 0.77 | 0.82 | 0.76 | 0.82 |
| $\varepsilon_{irr}$-$\varepsilon_{max}$ (%) | 0.34 | 0.34 | 0.44 | 0.45 |

IV. SUMMARY

Apparatuses are developed and used at the NHMFL to measure $I_C$ strain sensitivity in both longitudinal and transversal directions. These measurements are used to study the candidate Nb$_3$Sn strands for the series-connected hybrid magnet. In addition, ITER high $J_C$ strands were measured for $I_C$ irreversibility strain. The results show a consistent difference between strands of different designs.

ACKNOWLEDGMENT

Authors would like to acknowledge Dr. J.W. Ekin and Dr. N. Cheggour for the suggestions on the choice of backing plate material, and Dr. G. Nishijima for cross calibrations for the $I_C$ transversal stress measurements. We would also thank Nb$_3$Sn strand manufacturers OST, SMI, Luvata and Mitsubishi Electric for providing samples.

REFERENCES

[1] J. W. Ekin, "Strain scaling law for flux pinning in practical superconductors. Part 1: Basic relationship and application to Nb$_3$Sn conductors", *Cryogenics*, vol. 20, no. 11, pp. 611–624, 1980.
[2] J. W. Ekin, "Effect of transverse compressive stress on the critical current and upper critical field of Nb$_3$Sn", *J. Appl. Phys.*, vol. 62 no. 12, pp. 4829-4834, 1987.
[3] A Nijhuis and Y. Ilyin, "Transverse load optimization in Nb$_3$Sn CICC design; influence of cabling, void fraction and strand stiffness", *Supercond. Sci. Technol,*. vol. 19, pp. 945-962, 2006.
[4] Senkowicz B J, Takayasu M, Lee P J, Minervini J V and Larbalestier D C., "Relationship between architecture, filament breakage and critical current decay in Nb$_3$Sn composite wires repeatedly in-plane bent at room temperature", *IEEE Trans. Appl. Supercond.,* vol.15, pp. 3470, 2005.
[5] N. Mitchell, "Operating strain effects in Nb$_3$Sn cable-in-conduit conductors", *Supercond. Sci. Technol,*. vol. 18, pp. S396-S404, 2005.
[6] Y.H. Zhai and M.D. Bird, presentation 4J06 at MT-20.
[7] J.W. Ekin, "Strain scaling law and the prediction of uniaxial and bending strain effects in nultifilamentary superconductors", in *Proc. Topical Conf. on-A15-Supercond*, New York: Plenum, 1980, pp. 187
[8] J. Lu, K. Han, R.P. Walsh, and J.R. Miller, "Ic axial strain dependence of high current density Nb$_3$Sn conductors", *IEEE Trans. Appl. Supercond.*, vol. 17, no.2, pp. 2639, 2007.
[9] J. Lu, E.S. Choi, R.P. Walsh, R. Goddard, K. Han and J.R. Miller, "Characterization of Nb$_3$Sn Superconductors for Hybrid Magnets", *IEEE Trans. Appl. Supercond.*, vol. 17, no. 2, pp. 2651, 2007.
[10] L.T. Summers, M.W. Guinan, J.R. Miller, and P.A. Hahn, "A model for the prediction of Nb$_3$Sn critical current as a function of field, temperature, strain and radiation damage", *IEEE Trans. Mag.*, vol. 27, no. 2, pp. 2041-2044, 1991.
[11] A. Godeke, M. C. Jewell1, A. A. Golubov, B. Ten Haken and D. C. Larbalestier, "Inconsistencies between extrapolated and actual critical fields in Nb$_3$Sn wires as demonstrated by direct measurements of H$_{c2}$, H* and T$_C$", *Supercond. Sci. Technol.,* vol. 16, pp. 1019–1025, 2003.
[12] C. Senatore, D. Uglietti, V. Abacherli, A. Junod, and R. Flukiger, "Specific heat, a method to determine the $T_C$ distribution in industrial Nb$_3$Sn wires prepared by various techniques", *IEEE Trans. Appl. Supercond.*, vol. 17, no.2, pp. 2611, 2007.
[13] G. Nishijima, K. Watanabe, T. Araya, K. Katagiri, K. Kasaba, K. Miyoshi, "Effect of transverse compressive stress on internal reinforced Nb$_3$Sn superconducting wires and coils", *Cryogenics,* vol. 45, pp. 653–658, 2005.
[14] B. Seeber, A. Ferreira, V. Abächerli, T. Boutboul, L. Oberli, and R. Flükiger, "Transport Properties up to 1000 A of Nb$_3$Sn Wires Under Transverse Compressive Stress", *IEEE Trans. Appl. Supercond.*, vol. 17, no.2, pp. 2643, 2007.
[15] S. L. Bray, J. W. Ekin, R. Sesselmann, "Tensile measurement of modulus of elasticity of Nb$_3$Sn of room temperature and 4 K", *IEEE Trans. Appl. Supercond.*, vol. 7, no. 2, pp. 1451-1454, 1997.
[16] D. Ciazynski, "Review of Nb$_3$Sn conductors for ITER", in *24th symposium on fusion technology*. Sept 11-25, 2006, Warsaw, Poland.
[17] E. Barzi, T. Wokas, A. V. Zlobin, "Sensitivity of Nb3Sn Rutherford-type Cables to Transverse Pressure", *IEEE Trans. Appl. Supercond.*, vol. 15, No. 2, pp. 1541-1544, 2005.